\documentclass[conference]{IEEEtran}
\IEEEoverridecommandlockouts

\usepackage{cite}
\usepackage{amsmath,amssymb,amsfonts}
\usepackage{graphicx}
\usepackage{url}
\usepackage{tikz}
\usetikzlibrary{arrows.meta,positioning,calc,fit}

\def\RR{\mathbb{R}}

\def\EE{\mathbb{E}}

\def\PP{\text{Pr}}

\def\dKL{d_{\textsf{KL}}}

\def\piemp{\pi^{(\textsf{em})}}
\def\pitil{\pi^{(0)}}

\DeclareMathOperator{\Unif}{\mathsf{Unif}}

\newtheorem{theorem}{Theorem}[section]
\newtheorem{lemma}[theorem]{Lemma}
\newtheorem{corollary}[theorem]{Corollary}
\newtheorem{defn}{Definition}

\def\PPL{\textsf{PPL}}

\def\piopt{\hat{\pi}_{\textsf{opt}}}
\def\pitilopt{\hat{\pi}^{(0)}_{\textsf{opt}}}

\def\ntest{n^{\textsf{test}}}

\def\etil{\tilde{e}}

\DeclareMathOperator{\softmin}{\textsf{softmin}}

\begin{document}

\title{Semantic Smoothing for Language Models via Distribution Estimation and Embeddings}

\author{
\IEEEauthorblockN{Haricharan Balasundaram, Swathi Shree Narashiman, Pranay Mathur, Andrew Thangaraj}
\IEEEauthorblockA{Department of Electrical Engineering, IIT Madras\\
                   Chennai, India\\
                   Email: haricharanb,swathinarashiman,pranaym@smail.iitm.ac.in, andrew@ee.iitm.ac.in}
}

\maketitle

\begin{abstract}
We propose semantic smoothing, a smoothing method for language models that uses embeddings to share statistical observations across semantically similar contexts. The starting point is a decomposition of log-perplexity that motivates smoothing as a collection of distribution-estimation problems under Kullback-Leibler (KL) loss. We then show that, under a Lipschitz-logit model for embedding-based language generation, proximity of context embeddings implies proximity of the corresponding next-word distributions in KL divergence. Combining these observations, we formulate semantic smoothing as distribution estimation in KL loss with KL-proximity side information. For $n$ samples on a $d$-symbol alphabet with a side-information distribution at KL distance $\Delta$, we give an interpolation estimator with worst-case KL risk $O(\min\{\Delta,d/n\})$, and prove a matching-order lower bound for uniform side information. We extend the estimator to multiple and empirically estimated synonymous distributions. Experiments on synthetic Markov data and WikiText-103 bigram models using Word2Vec, GloVe, and GPT-2 embeddings show that semantic smoothing consistently reduces test perplexity when applied to add-constant and Kneser--Ney estimates.
\end{abstract}


\section{Introduction}
 
A $k$-gram language model posits a conditional distribution $\text{Pr}(w_k | w_1, w_2, \dots, w_{k - 1})$ for the next word $w_{k}$ given the $k - 1$ preceding words or context $w_1, w_2, \dots, w_{k - 1}$. The model is learned using a training corpus of sentences minimizing a metric called \textit{perplexity} (a modified version of log-loss that will be precisely defined later) and is expected to maintain low perplexity on a new test corpus of sentences. Since the test sentences will contain $k$-grams unseen in the training corpus, traditional Maximum Likelihood (ML) estimates result in unacceptably large test perplexity and \textit{smoothing}, which is the procedure for assigning suitable nonzero probabilities to unseen $k$-grams, becomes a very important requirement for language models \cite{chen_goodman}.

Smoothing methods for language models include i) add-constant smoothing studied by Laplace \cite{chen_goodman} and Lidstone \cite{lidstone}, and the interpolation/backoff-based ones proposed by ii) Katz \cite{katz}, iii) Jelinek-Mercer \cite{jelenik_mercer}, and iv) Kneser-Ney \cite{kneser_ney}. The interpolation/backoff-based methods use $l$-gram frequency counts for $l=1,\ldots,k$ and combine them using natural language heuristics. Kneser-Ney smoothing results in the lowest test perplexities in several data sets \cite{chen_goodman}. However, these models incur very high complexity as $k$ increases and, in recent years, they have been superseded in performance and popularity by large language models (LLMs) \cite{attention_is_all_you_need} that use embeddings of words over a large context to train a \textit{transformed} context embedding, which translates to a conditional next-word distribution estimate. It is well known that LLMs require very large data sets and significant compute power for training.

In this work, we propose a smoothing principle that uses semantic similarity, as measured by embeddings, to share probability mass across contexts. Classical smoothing methods share information across lower-order contexts or frequency classes. In contrast, semantic smoothing shares information across contexts whose embeddings are close, and therefore whose next-word distributions are expected to be close. We justify this principle in three steps.

\noindent 1) \textbf{Perplexity decomposition}: We observe that log-perplexity for a fixed context length decomposes into a model-independent, empirical conditional entropy of the test sequence plus a weighted sum of Kullback-Leibler (KL) divergences. So, minimization of empirical perplexity is achieved by minimizing the KL loss of distribution estimation after each context independently.

\noindent 2) \textbf{Word embeddings}: We determine semantic similarity of contexts through the proximity of their embeddings. We show that under suitable Lipschitz-logit conditions (satisfied by popular models such as Word2Vec \cite{word2vec,word2vec_2} and GPT-2 \cite{radford2018improving}) embedding proximity translates to a closeness of conditional next-word distributions in KL distance.

\noindent 3) \textbf{Semantic smoothing}: Combining the above two ideas, we propose semantic smoothing of language models. Semantically similar contexts (such as the synonyms `enormous' and `big' in bigram models) result in similar conditional distributions that can be combined to result in improved test perplexities.

The proposed semantic smoothing essentially reduces the problem of smoothing to a distribution estimation problem under KL loss given the knowledge that the unknown distribution is within a known KL distance of a side information distribution.

Our main theoretical result is as follows. For the underlying problem of distribution estimation with $n$ i.i.d. samples over a size-$d$ alphabet under KL loss with a known side information distribution at a proximity of $\Delta$ in KL distance, we provide a minimax upper bound of $O(\min(\Delta,d/n))$ through an \textit{interpolation} estimator that linearly interpolates between the add-constant estimator and the side information distribution with suitable weights. In the case where the side information distribution is uniform over the alphabet, we show a matching minimax lower bound using Assouad's lemma.

For experiments, we consider bigram semantic smoothing, where, for each context word $w$, we determine $m$ synonyms as the closest words $w_1,\ldots,w_m$ to $w$ in the embedding space and interpolate between the conditional distribution estimates $\text{Pr}_{\text{train}}(\cdot|w)$ and $\text{Pr}_{\text{train}}(\cdot|w_i)$, $i=1,\ldots,m$, with suitable weights as suggested by the theory. We evaluate the perplexity of our proposed bigram semantic smoothing method in the following scenarios: (1) synthetic Markov data with a low-rank factorization for the log of the transition matrix resulting in state embeddings, (2) WikiText-103 with Word2Vec/GloVe embeddings in a bigram model, and (3) WikiText-103 with GPT-2 embeddings in a bigram model of tokens\footnote{The code for the experiments can be found at \url{https://github.com/BaristaBandits/Distribution_Estimation}.}.

\noindent\textbf{Related work}: Outside of the classical smoothing methods mentioned earlier, recent work on $n$-gram smoothing includes Falahatgar et al.~\cite{falahatgar_20} that studies Kneser--Ney smoothing through competitive distribution estimation, and Malagutti et al.~\cite{malagutti2024role} that relates $n$-gram smoothing to regularization methods for neural language models. Our work is closest in spirit to this statistical line, but differs in that we formulate semantic smoothing as distribution estimation under KL loss with KL-proximity side information and prove minimax upper and lower bounds.

A second related direction is retrieval- and nearest-neighbor-based language modeling. The $k$NN-LM of Khandelwal et al.~\cite{khandelwal2020generalization} interpolates a neural language model with a distribution induced by nearest neighbors in representation space, and subsequent work studies why such nearest-neighbor language models improve perplexity~\cite{xu2023why}. These methods, like ours, use representation-space proximity to improve next-token prediction. However, our focus is different: rather than retrieving token-level neighbors at inference time, we use embedding proximity between contexts to construct semantically smoothed conditional distributions, with interpolation weights motivated by KL-risk bounds. 

Distribution estimation with side information is a framework introduced in~\cite{ISIT_paper} in the $\ell_2$--norm. The technical extensions in this work include KL loss with KL proximity (which has technical challenges especially in the lower bound), and the practical scenario where the synonymous distribution is to be estimated from the samples.

LLM alignment given a prompt is an important problem studied, for instance, in \cite{llm_alignment}, wherein the expected reward function has to be maximized subject to a KL--divergence constraint. This problem is mathematically similar to our distribution estimation with side information framework, with a difference that LLM alignment requires the final estimate to be at a distance $\Delta$ from a known distribution (the constraint) as opposed to $\Delta$ being the separation between the two synonymous distributions themselves.

\noindent\textbf{Organization of the Paper}: We introduce notation and state our results on perplexity decomposition, closeness of word embeddings, and i.i.d. distribution estimation with side information, respectively in Sections~\ref{sec:perplexity_decomposition},~\ref{sec:embeddings_proximity}, and~\ref{sec:semantic_smoothing}. We explain how we combine these results to get semantic smoothing for a language model in Section~\ref{sec:semantic_smoothing_practice} and provide experiments on synthetic and natural language data in Section~\ref{sec:experiments}. We provide concluding remarks in Section~\ref{sec:conclusions} and proofs for the theorems in Section~\ref{sec:proofs}. 

\section{Language Models: Perplexity Minimization and KL Loss}
\label{sec:perplexity_decomposition}

We consider a sequence of words $W^n=(W_1,W_2,\ldots,W_n)$ from natural language text to be a random process. The perplexity of a given word sequence $w^n=(w_1,\ldots,w_n)$, denoted $\PPL(w^n)$, is defined as 
\begin{equation}
    \PPL(w^n)=\exp\left(-\frac{1}{n}\log\text{Pr}(W^n=w^n)\right).
\end{equation}
The $\log$ function is to the base $e$ unless mentioned otherwise. This true perplexity is not computable because $\text{Pr}(W^n)$ is unknown and can only be estimated approximately. A popular approximation stems from the $k$-gram assumption.
\subsection{$k$-gram language models}
The one-sided $k$-gram context of the word $W_i$ is defined as the length-$(k{-}1)$ random vector $C_{i,k}(W^n) = (W_{i - k + 1}, W_{i - k + 2}, \dots, W_{i-1})$. $W_j$ for $j<1$ is assumed to be a special start-of-sentence word. Under the $k$-gram assumption, we suppose that the word sequence is a $(k{-}1)$-th order stationary Markov process, i.e. $W_i$ given the context $C_{i,k}(W^n)$ is conditionally independent of all other words. So, under $k$-gram, the probability of a given word sequence $w^n=(w_1,\ldots,w_n)$ is
\begin{align}
    \text{Pr}(W^n=w^n)=\prod_{i=1}^n \text{Pr}(W_i{=}w_i|C_i{=}c_i)=\prod_{i=1}^n p(w_i|c_i), \label{eq:tr}
\end{align}
where $C_i=C_{i,k}(W^n)$ and $c_i=c_{i,k}(w^n)$ are shorthand notations and $p(\cdot|c_i)$ denotes the transition probability of the $(k-1)$-th order Markov process for the context $c_i$. The perplexity of $w^n$ under the $k$-gram assumption with transition probability $p$, denoted $\PPL(p,w^n)$, is seen to be 
\begin{equation}
    \PPL(p,w^n)=\exp\left(-\frac{1}{n}\sum_{i=1}^n\log p(w_i|c_i)\right).
\end{equation}
A $k$-gram language model learns or estimates the transition probability $p(\cdot|c)$ for a $k$-gram context $c$ using a training corpus, which is a sequence of $S$ words that we will denote $u^S=(u_1,\ldots,u_S)$. For a particular language model $M$, we denote the estimated transition probability as $\hat{p}_{M,u^S}(\cdot|c)$ or simply $\hat{p}_M(\cdot|c)$, when the training corpus is not important.

\subsection{Test perplexity and its decomposition}
The figure of merit for a language model $M$ is its \textit{test} perplexity, i.e. $\PPL(\hat{p}_{M,u^S},v^T)$ of a test sequence of words $v^T=(v_1,\ldots,v_T)$, which is different from the train sequence $u^S$. A good model should have low test perplexity, which requires smoothing, i.e. assigning nonzero probabilities to contexts and $k$-grams unseen in the training corpus $u^S$ and likely to be seen in the test sequence $v^T$.

For a sequence $w^n$, a $k$-gram context $c$ and a word $w$, let 
\begin{align}
    p_{w^n,k}(c)&=\frac{1}{n}\sum_{i=1}^n I(c_{i,k}(w^n)=c) \text{ and }\nonumber\\
    p_{w^n,k}(w|c)&=\frac{\sum_{i=1}^n I(c_{i,k}(w^n)=c)I(w_i=w)}{\sum_{i=1}^n I(c_{i,k}(w^n)=c)},\nonumber
\end{align}
where $I(\cdot)$ is the indicator function and the second definition is valid when the denominator is nonzero. So, $p_{w^n,k}(\cdot)$ and $p_{w^n,k}(\cdot|c)$ are the empirical $k$-gram context distribution and empirical transition distribution given a $k$-gram context $c$, respectively. When $n$ and $k$ are clear, we use shorthands $p_w(\cdot)$ and $p_w(\cdot|c)$ to denote the empirical distributions.
\begin{theorem}
For a test sequence $v^T$ and a $k$-gram language model $M$, we have
\begin{multline}
\log(\PPL(\hat{p}_M, v^T)) = \sum_{c\text{ in }v^T} p_v(c)\,H(p_v(\cdot|c))\\ 
+ \sum_{c\text{ in }v^T} p_v(c)\, \dKL( p_v(\cdot | c) \,\|\, \hat{p}_M(\cdot | c)), \label{eq:perplexity_decomposition}
\end{multline}    
where $H(p)$ is the entropy of a distribution $p$ and $\dKL(p\,\|\,q)$ is the KL divergence between the distributions $p$ and $q$. Both the KL divergence and the entropy are in base $e$. 
\label{thm:perplexity_decomposition}
\end{theorem}
The decomposition in \eqref{eq:perplexity_decomposition} is a purely algebraic result for the perplexity of a fixed test sequence with the first empirical conditional entropy term being independent of the model $M$ (this term will be significant only for very short contexts), and the second conditional KL-divergence term measuring the closeness of the model's conditional distributions given a context to that of the corresponding empirical test distributions.

\subsection{Distribution estimation under KL Loss}
An important consequence of Theorem \ref{thm:perplexity_decomposition} is the following: for achieving low empirical perplexity, the quantity $\dKL( p_v(\cdot | c) \,\|\, \hat{p}_M(\cdot | c))$ is to be minimized and this minimization can be done independently for each context $c$. This leads to the classical problem of distribution estimation under KL loss using the training corpus $u^S$. However, since the test perplexity is with respect to a different test sequence $v^T$, a smoothing method becomes necessary. 

\section{Embeddings, Semantic Similarity and Proximity in KL Distance}
\label{sec:embeddings_proximity}

\subsection{Word and context embeddings}
\label{subsec:word_and_context_embeddings}
Modern LLMs embed words and contexts into a real vector space. A $t$-dimensional embedding $e$ of words from a vocabulary $V$ is a mapping taking $w\in V$ to $e_w\in\mathbb{R}^t$. Embeddings are learned such that closeness of $e_w$ and $e_{\tilde{w}}$ in the embedding space indicates semantic connections between the words $w$ and $\tilde{w}$. The context of a word $w_i$ occurring in the sequence $(w_1,w_2,\ldots)$ is either the one-sided $k$-gram context defined earlier, or the two-sided $k$-gram context $(w_{i-k+1},\ldots,w_{i-1},w_{i+1},\ldots,w_{i+k-1})$. Popular embeddings such as Word2Vec and GloVe can be understood as approximate, low-rank factorizations of the Pointwise Mutual Information (PMI) matrix derived from frequency counts of words occurring in contexts of a training corpus \cite{glove,word2vec}. 

For a context $c$ with $|c|$ words, a $t$-dimensional embedding $f$ is a mapping taking the embeddings of the words in the context $E_c=(e_w:w\in c)\in\RR^{t\times|c|}$ to a vector $f_c\in\RR^t$. Each column of $E_c$ is a length-$t$ embedding of a word in the context $c$. For example, transformers (the main components of LLMs) use the idea of \textit{attention} to transform $E_c$ (and positional encodings of words) in multiple computation steps into $(y_1,\ldots,y_{|c|})\in\RR^{t\times|c|}$, $y_i\in\RR^t$. The vector corresponding to the last context position $y_{|c|}$ is defined to be the embedding $f_c\in\mathbb{R}^t$ for a context $c$. 

In general, word and context embeddings are learned using training data and satisfy some semantic relationship heuristics. We will assume such embeddings are available and use them to impose Lipschitzness conditions on the natural language process.

\subsection{Embeddings to next-word probabilities: Lipschitzness}
In transformer-based LLMs, the context embedding $f_c$ is converted into \textit{logits} $e'_w f_c$, where $e_w\in\mathbb{R}^t$ is the embedding of a word $w\in V$ and $e'_w$ denotes the transpose of $e_w$. The logits are converted into a distribution over the words in $V$ using \textit{softmax}, which converts a vector $[a_1,a_2,\ldots]$ into a probability distribution $\text{Pr}(i)=\exp(a_i)/\sum_i\exp(a_i)$. 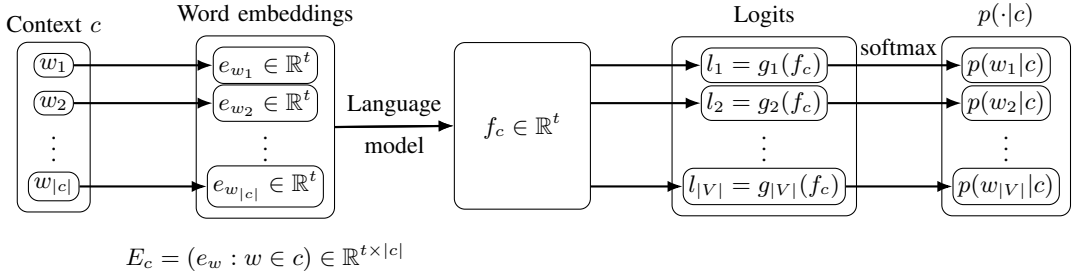
\begin{figure*}
    \centering
\begin{tikzpicture}[
    font=\small,
    >={Latex[length=2mm]},
    smallbox/.style={
        draw,
        rounded corners,
        inner sep=2pt,
        align=center
    },
    fcbox/.style={
        draw,
        rounded corners,
        inner sep=3pt,
        minimum width=1.8cm,
        minimum height=2.2cm,
        align=center
    },
    arr/.style={->, thick}
]

\node[smallbox] (w1) at (0,8mm) {$w_1$};
\node[smallbox] (w2) at (0,3mm) {$w_2$};
\node             (wd) at (0,-2mm) {$\vdots$};
\node[smallbox] (wn) at (0,-8mm) {$w_{|c|}$};

\node[
    draw,
    rounded corners,
    fit=(w1)(w2)(wd)(wn),
    inner sep=4pt,
    label=above:{Context $c$}
] (context) {};

\node[smallbox] (e1) at (28mm,8mm) {$e_{w_1}\in\mathbb{R}^t$};
\node[smallbox] (e2) at (28mm,3mm) {$e_{w_2}\in\mathbb{R}^t$};
\node             (ed) at (28mm,-2mm) {$\vdots$};
\node[smallbox] (en) at (28mm,-8mm) {$e_{w_{|c|}}\in\mathbb{R}^t$};

\node[
    draw,
    rounded corners,
    fit=(e1)(e2)(ed)(en),
    inner sep=4pt,
    label=above:{Word embeddings}
] (Ebox) {};

\node[below=2mm of Ebox] (Ec) {$E_c=(e_w:w\in c)\in\mathbb{R}^{t\times |c|}$};

\draw[arr] (w1.east) -- (e1.west);
\draw[arr] (w2.east) -- (e2.west);
\draw[arr] (wn.east) -- (en.west);

\node[fcbox] (fc) at (62mm,0mm) {$f_c\in\mathbb{R}^t$};

\draw[arr] (Ebox.east) -- node[above]{Language} (fc.west);
\draw[arr] (Ebox.east) -- node[below]{model} (fc.west);

\node[smallbox] (l1) at (94mm,8mm) {$l_1=g_1(f_c)$};
\node[smallbox] (l2) at (94mm,3mm) {$l_2=g_2(f_c)$};
\node             (ld) at (94mm,-2mm) {$\vdots$};
\node[smallbox] (lV) at (94mm,-8mm) {$l_{|V|}=g_{|V|}(f_c)$};

\node[
    draw,
    rounded corners,
    fit=(l1)(l2)(ld)(lV),
    inner sep=4pt,
    label=above:{Logits}
] (Lbox) {};

\coordinate (fc1) at ([yshift=8mm]fc.east);
\coordinate (fc2) at ([yshift=3mm]fc.east);
\coordinate (fcV) at ([yshift=-8mm]fc.east);

\draw[arr] (fc1) -- (l1.west);
\draw[arr] (fc2) -- (l2.west);
\draw[arr] (fcV) -- (lV.west);

\node[smallbox] (p1) at (126mm,8mm) {$p (w_1 | c)$};
\node[smallbox] (p2) at (126mm,3mm) {$p (w_2 | c)$};
\node             (pd) at (126mm,-2mm) {$\vdots$};
\node[smallbox] (pV) at (126mm,-8mm) {$p (w_{|V|} | c)$};

\node[
    draw,
    rounded corners,
    fit=(p1)(p2)(pd)(pV),
    inner sep=4pt,
    label=above:{$p(\cdot | c)$}
] (Pbox) {};

\draw[arr] (l1.east) -- node[above]{\ softmax} (p1.west);
\draw[arr] (l2.east) -- (p2.west);
\draw[arr] (lV.east) -- (pV.west);

\end{tikzpicture}
    \caption{Schematic depicting a language model - from embeddings to next-word probabilities.}
    \label{fig:schematic_embeddings}
\end{figure*}
Fig. \ref{fig:schematic_embeddings} shows the schematic of a language model that uses word and context embeddings that are converted to next-word probabilities using logit-softmax.

We consider the following generalized Lipschitz-logit model for the natural language process that generates the training corpus $u^S$ and the test sequence $v^T$:
\begin{defn}[$(L, \epsilon)$-Lipschitz-logit process]
A random process on words in a vocabulary $V$ with a context defined for each word and a transition probability $\text{Pr}(\cdot|c)$ over $V$ for a context $c$ is said to be $(L, \epsilon)$-Lipschitz-logit if
\begin{equation}
    \PP(w | c) = \frac{\exp(g_w(f_c) + \epsilon_{c, w})}{\sum_{y \in V} \exp(g_{y}(f_c) + \epsilon_{c, y})}, \quad |\epsilon_{c, w}| \le \epsilon \, \forall (w, c), \label{eqn:softmax_probab}
\end{equation}
where $f_c\in\RR^t$ is the embedding for the context $c$, the function $g_w: \RR^t \to \RR$ is $L$-Lipschitz satisfying $|g_w(f_c)-g_w(f_{\tilde{c}})|\le L\|f_{c}-f_{\tilde{c}}\|_2$, and $\epsilon_{c, w}$ is the permitted error.
\end{defn}
As pointed out earlier, text generated by transformer-based LLMs clearly satisfy the Lipschitz-logit condition with $g_w(f_c)=e'_wf_c$ and $L=\max_w\|e_w\|_2$.

Though our proposed smoothing method is motivated by the Lipschitz-logit assumption, we do not require any knowledge or use of the functions $g_w$, which may not be available or may be computationally intensive to evaluate. Compared to the number of parameters in $g_w$ of modern LLMs, the size of the embeddings is significantly smaller.

\subsection{KL proximity}
For a Lipschitz-logit process, as shown in the theorem below, closeness in context embeddings results in closeness in the KL--divergences between the corresponding conditional probability distributions.
\begin{theorem}
\label{thm:embeddings_lipschitz}
In an $(L, \epsilon)$-Lipschitz-logit process, let $c, \tilde{c}$ be two contexts and let their embeddings be $f_{c}$ and $f_{\tilde{c}}$. Then, 
\begin{equation}
    \dKL (\PP(\cdot | c) \| \PP(\cdot | \tilde{c})) \le 2 L \|f_{c} - f_{\tilde{c}}\|_2 + 4 \epsilon.
\end{equation}
\end{theorem}
We have chosen the 2-norm in the statement of the above theorem for simplicity. A similar result will hold for all other equivalent norms. 

In several practical examples, the embedding function $f:\RR^{t\times|c|}\to\RR^t$ can be shown to be Lipschitz, i.e. $\|f_c-f_{\tilde{c}}\|_2\le L_f\|E_c-E_{\tilde{c}}\|_F$, where $\|\cdot\|_F$ is the Frobenius norm. In Appendix~\ref{appendix:embeddings}, we prove this for the Word2Vec Continuous-Bag-Of-Words (CBOW) model, transformer-based models (with bounded input norms) and GloVe-based bigram models. In such cases, the KL proximity of the next-word probabilities can be expressed in terms of word embeddings. This is shown in the following corollary to Theorem \ref{thm:embeddings_lipschitz}.
\begin{corollary}
If the context embedding function $f_c$ is $L_f$-Lipschitz as defined above, then
\begin{equation}
   \dKL (\PP(\cdot | c) \| \PP(\cdot | \tilde{c})) \le 2 L L_f \|E_c - E_{\tilde{c}}\|_F + 4 \epsilon
\end{equation}
for an $(L,\epsilon)$-Lipschitz-logit process.
\label{thm:embedding_mismatch}
\end{corollary}
The results in this and the previous section firmly establish the theoretical basis of semantic smoothing, i.e. estimation of the conditional distribution given a context under the metric of KL loss with the knowledge that the conditional distributions of semantically similar contexts are proximal in KL distance. Next, we consider such distribution estimation problems and derive minimax results for them.

\section{Distribution Estimation under KL Loss with KL Proximity Side Information}
\label{sec:semantic_smoothing}

Consider $n$ i.i.d. samples $X^n = (X_1, X_2,\dots, X_n)$ from a distribution $\pi = (\pi_1, \pi_2, \dots, \pi_d)$. Let $N_i = \sum_{j \in [n]} I(X_j = i)$ denote the individual letter counts. Let $\piemp_i = \frac{N_i}{n}$ denote the empirical distribution.

For the distribution estimation problem of estimating $\pi$ from the samples $X^n$ under expected KL loss, minimax results have been studied in \cite{cover_dist_est,KT_dist_est,catoni_laplace_est,kl_div_estimation_1,kl_div_estimation_2} and a minimax optimal estimator is the variable add-constant estimator $\piopt$ that assigns $\piopt(i)\propto r+\beta_{r}$ to the letter $i$ when $N_i=r$ and the constant $\beta_r$ varies with $r$ as specified in \cite{kl_div_estimation_3}. The result is captured in the theorem below.
\begin{theorem}
\begin{multline}
    \min_{\hat \pi} \max_{\pi} \EE[\dKL(\pi \| \hat \pi(X^n))]\nonumber
    \\ = \max_{\pi} \EE[\dKL(\pi \| \piopt(X^n))] = \frac{d - 1}{2n} + o \left(\frac{1}{n} \right),
\end{multline}
where $\hat{\pi}$ is an estimator for the distribution $\pi$ and the expectation is over the samples $X^n\sim$ i.i.d. $\pi$.
\label{thm:dist_est_upper_bound_optimal}
\end{theorem}

\subsection{Minimax distribution estimation with side information}
We now consider the scenario where, along with the i.i.d. samples $X^n\sim\pi$, we have side information that $\pi$ is \textit{close} to another distribution $\pitil
 = (\pitil_1, \pitil_2, \dots, \pitil_{d_0})$. More precisely, our side information is that the distribution $\pi\in B_{\text{KL}}(\pitil, \Delta) = \{\pi': \dKL(\pi' \| \pitil) \le \Delta\}$, which is the $\Delta$-neighborhood around $\pitil$ in KL distance. We assume supports of $\pi$ and $\pi^{(0)}$ are such that $\Delta$ is finite.  In terms of language models, $\pi=\text{Pr}(\cdot|c)$, $\pitil=\text{Pr}(\cdot|\tilde{c})$ and $\Delta$ is the semantic similarity between the two contexts $c$ and $\tilde{c}$ evaluated through the Lipschitz-logit assumption. We will call $\pi$ and $\pitil$ as \textit{synonymous} distributions. For now, we assume that $\pi^{(0)}$ is known exactly, and we will relax this assumption later.

The minimax distribution estimation problem with side information can be stated as
\begin{equation}
    \min_{\hat{\pi}}\max_{\pi\in B_{\text{KL}}(\pitil,\Delta)}\EE[\dKL(\pi \| \hat{\pi}(X^n))],
    \label{eq:newminimax}
\end{equation}
where the expectation is over the samples $X^n\sim$ i.i.d. $\pi$. This specific problem with expected KL loss and KL proximity has not been considered in earlier work. 

Our first result for the minimax problem in \eqref{eq:newminimax} is the following upper bound.
\begin{theorem}
    \label{thm:dist_est_upper_bound}
    The linear interpolation estimator $\hat \pi = \alpha \piopt(X^n) + (1 - \alpha) \pitil$ for appropriate values of $\alpha\in[0,1]$ achieves 
    \begin{equation}
        \max_{\pi \in B_{\text{KL}}(\pitil, \Delta)} \EE[\dKL (\pi \| \hat \pi)] \le \min \left( \frac{d - 1}{2n} + o \left(\frac{1}{n} \right), \Delta \right),
    \end{equation}
    where the expectation is over the samples $X^n\sim$ i.i.d. $\pi$.
\end{theorem}

Our second result is a matching lower bound for a special case of $\pitil$.
\begin{theorem}
\label{thm:dist_est_lower_bound}
Let $\pitil = \Unif[d]$ be the uniform distribution on an alphabet of size $d$. Then,
\begin{equation}
    \min_{\hat \pi} \max_{\pi \in B_{\text{KL}}(\pitil, \Delta)} \EE[\dKL(\pi \| \hat \pi)] \ge \frac {1}{32} \min \left(\frac{d}{4 n}, \frac 1{4}, \Delta \right) e^{-4}.
\end{equation}
\end{theorem}
For $\Delta<1/4$, it can be seen that the lower bound for uniform $\pitil$ and upper bound are both $\Theta \left(\min \left(\Delta, \frac dn \right) \right)$.

\subsection{Extension to $m$ synonyms}
\label{sec:m-synonym-extension}
The results above are stated for one synonymous distribution $\pitil$ with $\Delta$. It is easy to see that if there are $m$ synonymous distributions $\pi^{(1)}, \pi^{(2)} \dots \pi^{(m)}$ with respective KL proximities $\Delta_1, \Delta_2, \dots, \Delta_m$, the linear interpolation estimator 
\begin{equation}
        \alpha_0 \piopt + \alpha_1 \pi^{(1)} + \alpha_2 \pi^{(2)} + \dots + \alpha_m \pi^{(m)}
        \label{eqn:interp_multiple_synonyms}
\end{equation}
with $\alpha_i\ge0$, $\sum_{i = 0}^m \alpha_i = 1$ achieves the upper bound of
\begin{equation}
        O \left( \min \left(\frac {d - 1}{2n}, \Delta_1, \Delta_2, \dots, \Delta_m \right) \right).
\end{equation}
The bound above is minimized by placing all mass on the source with the smallest loss. In practice, we use a softmin (or other similar functions) over the estimated loss proxies to obtain a more stable interpolation as follows:
\begin{equation}
\alpha_i \propto \begin{cases}
    \exp(-\tau(\frac {d - 1}{2n})), & i = 0, \\
    \exp(-\tau\Delta_i), & \text{otherwise},
    \end{cases}
\end{equation}
where $\tau$ is a temperature parameter.

\subsection{Estimated synonymous distribution}
\label{sec:final_interpolation}
So far, we have assumed that the synonymous distribution $\pitil$ is known completely. However, in the context of language models, the distributions $\pi$ and $\pitil$ arise as distributions $\text{Pr}(\cdot|c)$ and $\text{Pr}(\cdot|\tilde{c})$ following similar contexts $c$ and $\tilde{c}$. So, in practice, we will not directly know $\pitil$, but we will have, say, $n_0$ samples from $\pitil$, where $n_0$ is the number of times the context $\tilde{c}$ occurs in the training sample, and we have to estimate the distribution $\pitil$ from the $n_0$ i.i.d. samples. 

In this version of the problem of estimating $\pi$, we have $n$ samples $X^n\sim$ i.i.d. $\pi=(\pi_1,\ldots,\pi_d)$, $n_0$ samples $Y^{n_0}\sim$ i.i.d. $\pitil=(\pitil_1,\ldots,\pitil_{d_0})$ with the side information that $\pi \in B_{\text{KL}}(\pitil, \Delta)$. We proceed with a \textit{plugin} approach as follows. First, from the $n_0$ samples $Y^n$, we find the variable add-constant estimate $\pitilopt(Y^n)$ for $\pitil$, which is minimax optimal as claimed in Theorem~\ref{thm:dist_est_upper_bound_optimal}. Next, we use $\pitilopt(Y^n)$ in place of $\pitil$ in the linear interpolation estimator in Theorem \ref{thm:dist_est_upper_bound} to obtain an estimator for $\pi$. For this plugin-interpolated estimator we prove a risk upper bound.
\begin{theorem}
    The estimator 
    $$\hat \pi = \alpha \piopt(X^n) + (1 - \alpha) \pitilopt(Y^n)$$ 
    for appropriate values of $\alpha\in[0,1]$ satisfies
    \begin{multline}
        \EE[\dKL (\pi \| \hat \pi)] \le \min \Bigg(\frac{d - 1}{2 n} + o \left(\frac{d}{n} \right), \\ 
        \Delta + \log \left(1 + \frac{d_0}{n_0} \right) + \log \left(2\right) \Bigg).\nonumber
    \end{multline}
\label{thm:dist_est_upper_bound_synonym_sample}
\end{theorem}
For typical parameters, if the approximation $\log(1 + x) \approx x$ holds, we get a rough bound of the order of $\min \left(\frac{d}{n},\Delta+\frac{d_0}{n_0} \right)$. As before, the estimator and bound of Theorem \ref{thm:dist_est_upper_bound_synonym_sample} can be extended analogously from a single synonym to $m$ synonyms.

\section{Semantic Smoothing: Theory to Practice}
\label{sec:semantic_smoothing_practice}

Consider a corpus with training sequence $u^S=(u_1,\ldots,u_S)$ and test sequence $v^T=(v_1,\ldots,v_T)$, which we will assume are generated through a common random process $\mathcal{P}$ that is $(L,\epsilon)$-Lipschitz-logit with next-word probabilities $p_{\mathcal{P}}(\cdot|c)$ for a context $c$, embeddings $e_w$ for a word $w$ and $f_c$ for the context $c$. The embeddings are assumed to be available along with $u^S$ for training a language model $M$ with parameters $\theta(M)$ and obtain the next-word probabilities $\hat{p}_{M,u^S}(\cdot|c)$ so as to minimize the test perplexity $\PPL(\hat{p}_{M,u^S},v^T)$.

Theoretically, we assume random $U^S\sim\mathcal{P}$, $V^T\sim\mathcal{P}$ and consider the following problem of minimizing the expected log perplexity:
\begin{equation}
   \min_{\theta(M)} \EE_{U^S,V^T}[\log(\PPL(\hat{p}_{M,U^S},V^T)].
\end{equation}
Using the decomposition of perplexity in Theorem \ref{thm:perplexity_decomposition} (with suitable separability assumptions for context stationary probabilities) and the proximity results of Theorem \ref{thm:embeddings_lipschitz} (and its corollaries), we relate the problem to the following distribution estimation with side information problem for each context $c$:
\begin{equation}
    \min_{\theta(M)} \EE_{U^S}[\dKL(p_{\mathcal{P}}(\cdot|c)\,\|\,p_{M,U^S}(\cdot|c)]
\end{equation}
with the side information that $p_{\mathcal{P}}(\cdot|c)$ is close in KL distance to $p_{\mathcal{P}}(\cdot|c_i)$, $i=1,\ldots,m$, where the $c_i$ are neighboring or synonymous contexts in the embedding space. 

For a specific training sequence $U^S=u^S$, the language model $M$ estimates $p_{\mathcal{P}}(\cdot|c)$ as $\hat{p}_{M,u^S}(\cdot|c)$. Using the ideas of Section \ref{sec:final_interpolation} and, in particular, Theorem \ref{thm:dist_est_upper_bound_synonym_sample}, we define the semantically smoothed version of the language model's estimate, denoted $\hat{p}^{\text{sem}}_{M,u^S}(\cdot|c)$, as follows:
\begin{align}
    \hat{p}^{\text{sem}}_{M,u^S}(\cdot|c) &= \alpha^{(c)}_c\, \hat{p}_{M,u^S}(\cdot|c)\nonumber\\
    &+ \alpha^{(c)}_{c_1}\, \hat{p}_{M,u^S}(\cdot|c_1) + \cdots + \alpha^{(c)}_{c_m}\, \hat{p}_{M,u^S}(\cdot|c_m),\label{eqn:semantic_smoothing_interpolation}
\end{align}
where $\alpha^{(c)}_c,\alpha^{(c)}_{c_i}\in[0,1]$, $i=1,\ldots,m$ and $\alpha^{(c)}_c+\sum_{i=1}^m \alpha^{(c)}_{c_i}=1$ are linear interpolation coefficients that are treated as hyperparameters and chosen using ideas motivated by the theoretical bounds.

Suppose the context $c$ occurs $n_c$ times in $u^S$, and the context $c_i$ appears $n_{c_i}$ times. To use the bounds of Theorem \ref{thm:dist_est_upper_bound_synonym_sample}, we require knowledge of the alphabet size $d_c$ of words that follow $c$ and the alphabet size $d_{c_i}$ of words that follow $c_i$. In experiments, when the sizes are unknown, we estimate them using the samples of next words following the respective contexts in $u^S$. Next, we estimate the KL proximity of $c_i$ to $c$ using Theorem \ref{thm:embeddings_lipschitz} as 
\begin{equation}
    \Delta_{cc_i}=L\|f_c-f_{c_i}\|
\end{equation}
for a suitable norm (usually $1$ or $2$ norm) and a hyperparameter Lipschitz constant $L$ chosen suitably. 

Applying the ideas in Section \ref{sec:m-synonym-extension}, we pick $\alpha^{(c)}_{c'}$ as being proportional to a function of the respective KL loss proxies (upper bounds from Theorem \ref{thm:dist_est_upper_bound_synonym_sample}) as follows (this requires a mild generalization of the proof of Theorem \ref{thm:dist_est_upper_bound_synonym_sample} for an arbitrary $\beta$):
\begin{align}
    \alpha^{(c)}_{c'}\propto \begin{cases}
        \phi\left(\frac{d_c-1}{2n_c}\right),&c'=c,\\
        \phi\left(\Delta_{cc'}+\log\left(1+\frac{\beta d_{c'}}{n_{c'}} \right) + [\log(\frac 1\beta)]_+\right),&c'=c_i,
    \end{cases}
    \label{eqn:lambda_value}
\end{align}
where $(x)_+$ denotes $\max(x, 0)$ and the function $\phi(x)$ could be chosen as $\exp(-\tau x)$ for $\softmin$ or another similar one. For any language model $M$, we will use the loss proxies of an add-$\beta$ as shown in \eqref{eqn:lambda_value}. 

For any language model $M$, the above idea of semantic smoothing can be applied to obtain new estimates of next-word probabilities. Theoretically, this approach is motivated by the minimization of expected test perplexity under the assumptions described in the sections above. To verify the theory, we provide experimental results with both a synthetic and a natural language data set. 

\section{Experiments}

\label{sec:experiments}

\subsection{Synthetic experiments on Markov data}

We assume a bigram model and generate the train and test sequences by a Markov process with transition matrix $P_{cw} = \PP(w | c)$, where $c$ denotes the context which is the previous word and $w$ denotes the word following the context. The Markov process has $|V| = 100$ states and the transition matrix has $10$ distinct rows repeated $10$ times each where the distinct rows are randomly generated. 

The embeddings are also $t = 10$ dimensional vectors.
The embeddings are obtained by factorizing the matrix $Q_{cw} = \log\left(\frac{\PP(w | c)}{\PP(c)} \right)$ as $Q_{cw} = e'_c \etil_w$ ($e'$ is the transpose of $e$). Such a factorization exists since $P$ is low rank. So,
\begin{gather}
    \exp(e'_c \etil_w) = \frac{\PP(w | c)}{\PP(c)}, \quad \sum_{y \in V_c} \exp(e'_c \etil_y) = \frac{1}{P(c)},\nonumber\\
    \PP(w | c) = \frac{\exp(e'_c \etil_w)}{\sum_{y \in V_c} \exp(e'_c \etil_y)},\nonumber
\end{gather}
which is $(L,0)$-Lipschitz-logit with $g_w(x) = x' \etil_w$ and $L = \max_{w \in V} \|\etil_w\|_2$.

\begin{figure}[!ht]
    \centering
    \includegraphics{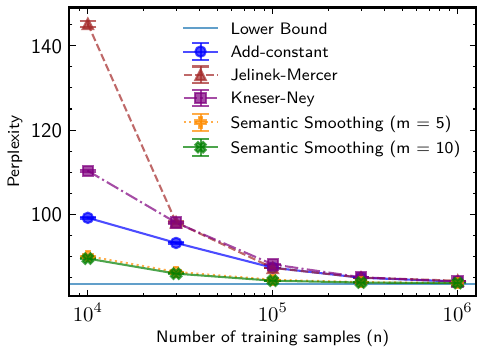}
    \caption{Perplexity vs Number of Training samples for semantic smoothing with varying number of synonyms.}
    \label{fig:synthetic_simulations}
\end{figure}

The add-$\beta$ model uses
\begin{equation}
\hat p_{\text{add}-\beta}(w|c) = \frac{N_{c, w} + \beta}{N_{c} + \beta|V|},  
\end{equation}
where $N_{c, w}$ is the number of times the word $w$ occurs after the context $c$ in the training data, $N_c = \sum_{w \in V} N_{c, w}$ is the number of occurrences of the unigram $c$ in the training data, and $\beta>0$ is a constant.

We consider the add-$\beta$ estimator with $\beta = 1$ and its semantically smoothed versions with $m \in \{5, 10\}$ synonyms that are the words closest to the context word in the embedding space. The softmin function with $\tau = 1$ is used for interpolation. We take $\ntest = 5\times10^6$ samples for computing test perplexity with $10$ repetitions for obtaining error bars. A plot of test perplexity vs the number of training samples $n$ is shown in Fig.~\ref{fig:synthetic_simulations}. 

It is seen that the test perplexity improves significantly by semantic smoothing. The semantic-smoothed estimators perform better than the naive add-constant, Kneser-Ney and Jelinek-Mercer estimators, and come close to the conditional entropy lower bound $H(p(\text{word}|\text{context}))$ from Theorem~\ref{thm:perplexity_decomposition} even for small values of $n$. 

\subsection{Experiments on natural language data}

We evaluate the semantic smoothing method on natural language data to show its effectiveness even when there is no simple bigram process for data generation. Our goal is to merely show the effectiveness of semantic smoothing and no effort has been made to optimize for the lowest perplexities by hyperparameter tuning.

\noindent\textbf{Dataset, embeddings and preprocessing}: We run experiments on the WikiText-103 dataset \cite{merity2016pointer}, using the inbuilt train-test split as available in the Python Datasets \cite{lhoest2021datasets} module. The validation data is not used. A token (typically a word) is defined to be the lowest level of division of a sentence. The data is first preprocessed by i) making letters lowercase, ii) replacing numbers with a \texttt{<num>} token and URLs with a \texttt{<url>} token, iii) removing references, non-ASCII characters and punctuation, and iv) normalizing whitespaces.

For the simulations, we use Word2Vec ($300$ dimensions) \cite{word2vec}, GloVe ($100$ dimensions) \cite{glove}, and GPT-2 ($768$ dimensions) \cite{radford2018improving} embeddings. In the case of Word2Vec and GloVe embeddings, tokens are created by i) splitting the sentences at whitespaces, and ii) padding the beginning and end of each sentence with \texttt{<bos>} and \texttt{<eos>} tokens, respectively. For GPT-2 embeddings, we use the GPT-2 tokenizer \cite{radford2018improving}. We construct a corpus consisting of $100,000$ training sentences and $10,000$ test sentences from Wikitext-103 after preprocessing and sentence tokenization.

Let $V$ be the vocabulary, defined formally to be the set of unique tokens appearing in either the train or the test sequence. Using standard word-level tokenization, the vocabulary size is found to be $66,753$. For GPT2 tokenization, the preprocessing pipeline produces a reduced vocabulary size of $27,007$, due to byte-pair encoding. 

\noindent\textbf{Language models}: We again consider bigram language models, such as the add-$\beta$ estimator defined earlier and the Kneser-Ney (KN) model. The KN model with an absolute discount $D$, for a context $c$ which is the 
previous word and the next word $w$, is defined as
\[
\hat p_{\text{KN}-D}(w|c) =
\frac{\max(N_{c, w} {-} D,0)}{N_{c}} + \lambda_{c} P_{\text{cont}}(c),
\]
where $\lambda_{w}=D \, N_1^{+}(w, \cdot)/N_w$, $P_{\text{cont}}(w) =
N_1^{+}(\cdot, w)/N_1^{+}(\cdot, \cdot)$, $N_1^{+}(w, \cdot)$ is the number of unique continuations after $w$, $N_1^{+}(\cdot, w)$ is the number of unique contexts before $w$, and $N_1^{+}(\cdot, \cdot)$ is the total number of unique bigrams.

\noindent\textbf{Semantic smoothing}: For a bigram model, the context $c$ is a word $w$ and the embedding $f_c$ equals the word embedding $e_w$. We compute the KL proximity as $\Delta_{w\tilde{w}}=5\|e_w-e_{\tilde{w}}\|_1$, where we have chosen the Lipschitz constant as $5$ based on the maximum norm of the word embeddings. For each $w \in V$, we estimate the alphabet size of the words following $w$, $\hat{d}_w$, using the Chebyshev polynomial estimator from \cite{wu2015chebyshev}. We pick the $m$ synonymous words of a given word $w$ to be the $m$ words $\tilde{w}$ with the smallest values of $\Delta_{cc'}+\log\left(1+\frac{\beta \hat{d}_{c'}}{n_{c'}} \right)$, choosing $\beta = 0.005$. The interpolation function $\phi$ is set to be $\phi(x)=1/x$. For any word in the train or test sequence that does not have an embedding, we skip semantic smoothing.

\begin{figure}[!ht]
    \centering
    \includegraphics{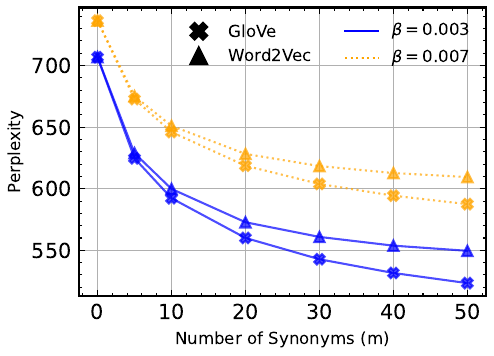}
    \caption{Perplexity vs no. of synonyms for add-$\beta$ model. The conditional entropy lower bound from Theorem~\ref{thm:perplexity_decomposition} is $33.51$.}
    \label{fig:add-k}
\end{figure}
For the add-constant model, a plot of test perplexity versus number of synonyms of semantic smoothing using GloVe and Word2Vec embeddings is shown in Fig.~\ref{fig:add-k}. We can see that the test perplexity drops as the number of synonyms increases for different constants $\beta$.

\begin{figure}[ht]
    \centering
    \includegraphics{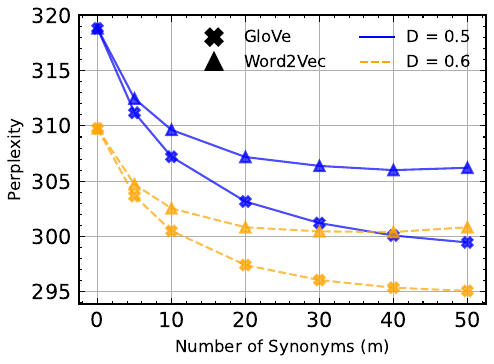}
    \caption{Perplexity vs no. of synonyms for Kneser-Ney model. The conditional entropy lower bound from Theorem~\ref{thm:perplexity_decomposition} is $33.51$.}
    \label{fig:kn}
\end{figure}
A similar plot for the Kneser-Ney model is shown in Fig. \ref{fig:kn}, and an improvement in perplexity is observed with increasing $m$ for different values of the absolute discount $D$. Comparing Figs. \ref{fig:add-k} and \ref{fig:kn}, we observe that the absolute perplexities are significantly better for the KN model and that semantic smoothing ($m=50$) is effective in both cases in different proportions -- about $700$ to $520$ for add-$0.003$ model and about $310$ to $295$ for KN-$0.6$.

\begin{figure}[ht]
    \centering
    \includegraphics{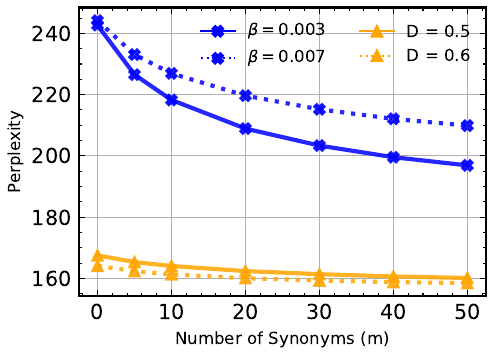}
    \caption{Perplexity vs no. of synonyms for add-$\beta$ and  KN-$D$ models using GPT-2 embeddings.}
    \label{fig:kn1}
\end{figure}

Finally, in Fig. \ref{fig:kn1}, we plot perplexity versus number of synonyms when using GPT-2 embeddings and tokenization, which results in a significantly lower absolute perplexity value. However, semantic smoothing ($m=50$) results in improvements -- about $240$ to under $200$ for add-$0.003$ and about $164$ to $160$ for KN-$0.6$.

In summary, while several improvements are potentially possible, these initial experiments with both synthetic and natural language data establish the effectiveness of semantic smoothing for reducing the perplexity metric in practice. 

\section{Concluding remarks}
\label{sec:conclusions}

We proposed semantic smoothing, a framework for improving language-model probability estimates by combining information across semantically similar contexts. The method has information-theoretic backing through a distribution estimation with side information problem. The theoretical and experimental results suggest that embedding geometry can provide useful smoothing information beyond frequency counts alone.

Several directions remain open. First, the present experiments focus on bigram models; extending the method to longer contexts requires scalable nearest-neighbor search over context embeddings and careful estimation of the corresponding interpolation weights. Second, the current weighting rules are motivated by minimax upper bounds but are not optimized end-to-end for held-out perplexity; data-dependent or adaptive weighting may further improve performance. Third, it will be interesting to train/use embeddings optimized for the purpose of semantic smoothing as opposed to using general purpose ones. Finally, the distribution-estimation problem with side information appears to be of independent interest, particularly when the side-information distribution is itself estimated from samples. Several aspects of this problem including optimal minimax rates for a general side information distribution remain open.

\section{Proofs of Theorems}
\label{sec:proofs}

\subsection{Proof of Theorem~\ref{thm:perplexity_decomposition}}

Let $N_{c, w} = \sum_{i=1}^T I(c_{i,k}(v^T)=c)I(v_i=w)$, i.e. the number of times $w$ occurs after a $k$-gram context $c$ in $v^T$. Note that
\begin{IEEEeqnarray}{rCl}
    \IEEEeqnarraymulticol{3}{l}{\log(\PPL(\hat{p}_M, v^T))} \nonumber \\
    &=& -\frac{1}{T} \sum_{i = 1}^T \log(\hat p_M(v_i | c_i)) \nonumber \\
    & \overset{(a)}{=} & \frac{-1}{T} \sum_{(c, w)} N_{c, w} \log(\hat p_M(w | c)) \nonumber\\
    & \overset{(b)}{=} & - \sum_{(c, w)} p_v(c) p_v(w | c) \log \left(p_v(w | c) \cdot \frac{\hat p_M(w | c)}{p_v(w | c)} \right)  \nonumber\\
    & = & \sum_{(c, w)} p_v(c) p_v(w | c) \left[\log \left( \frac{1}{p_v(w | c)} \right) + \log \left(\frac{p_v(w | c)}{\hat p_M(w | c)} \right) \right] \nonumber \\
    &=&  \sum_c p_v(c)\,H(p_v(\cdot|c)) + \sum_{c} p_v(c)\, \dKL( p_v(\cdot | c) \,\|\, \hat{p}_M(\cdot | c)) \nonumber,
\end{IEEEeqnarray}
where $(a)$ follows by regrouping terms in the summation to be over all word-context tuples in the test sequence, $(b)$ follows from $N_{cw} = p_v(c) p_v(w | c) \, T$ by definition.

\subsection{Proof of Theorem~\ref{thm:embeddings_lipschitz}}
Let 
\begin{align*}
Z &= \sum_{y \in V} \exp(g_y(f_{c}) + \epsilon_{c,y})),\\
\tilde{Z} &= \sum_{y \in V} \exp(g_y(f_{\tilde{c}}) + \epsilon_{\tilde{c},y}))    
\end{align*}
denote the denominators of the next-word probability expression in \eqref{eqn:softmax_probab}. Now, 
\begin{IEEEeqnarray}{rCl}
    \log \frac{\tilde{Z}}{Z} &=& \log \frac{\sum_{y \in V} \exp(g_y(f_{\tilde{c}}) + \epsilon_{\tilde{c},y})}{\sum_{y \in V} \exp(g_y(f_{c}) + \epsilon_{c,y})} \nonumber\\
    &\le& \log \frac{\sum_{y \in V} \exp(g_y(f_{c}) + L \|f_{c} - f_{\tilde{c}}\| + 2 \epsilon + \epsilon_{c,y})}{\sum_{y \in V} \exp(g_y(f_{c}) + \epsilon_{c,y})} \nonumber \\
    &=& L \|f_{c} - f_{\tilde{c}}\| + 2 \epsilon.
\end{IEEEeqnarray}
Next, note that
\begin{IEEEeqnarray}{rCl}
    \log \left(\frac{\PP(x | c)}{\PP(x | \tilde{c})}\right) &=& g_x(f_{c}) + \epsilon_{c,x} - g_x(f_{\tilde{c}}) - \epsilon_{\tilde{c},x} - \log \left(\frac{Z}{\tilde{Z}} \right) \nonumber \\
    &\le& 2 \cdot L \|f_{c} - f_{\tilde{c}}\| + 4 \epsilon.
\end{IEEEeqnarray}
From this,
\begin{IEEEeqnarray}{rCl}
    \IEEEeqnarraymulticol{3}{l}{\dKL (\PP(\cdot | c) \| \PP(\cdot | \tilde{c}))} \nonumber\\
    &=& \sum_{x \in V} \PP(x | c) \log \left(\frac{\PP(x | c)}{\PP(x | \tilde{c})}\right) \nonumber\\
    &\le& \left(\sum_{x \in V} \PP(x | c) \right) \cdot \left(2 \cdot L \|f_{c} - f_{\tilde{c}}\| + 4 \epsilon \right) \nonumber\\
    &=& 2 \cdot L \|f_{c} - f_{\tilde{c}}\| + 4 \epsilon.
\end{IEEEeqnarray}

\subsection{Proof of Corollary~\ref{thm:embedding_mismatch}}
The proof follows directly from Theorem~\ref{thm:embeddings_lipschitz} and the $L_f$-Lipschitzness.

\subsection{Proof of Theorem~\ref{thm:dist_est_upper_bound}}

Note from \cite[Theorem 2.7.2]{elements_of_information_theory} that the KL--divergence is convex in the second argument. From this, we have
\begin{IEEEeqnarray}{rCl}
    \IEEEeqnarraymulticol{3}{l}{\max_{\pi \in B_{\text{KL}}(\pitil, \Delta)} \EE[\dKL (\pi \| \hat \pi)]} \nonumber \\
    &\le& \max_{\pi \in B_{\text{KL}}(\pitil, \Delta)} \left(\alpha \EE[\dKL (\pi \| \piopt)] + (1 - \alpha) \dKL (\pi \| \pitil) \right) \nonumber \\
    &\le& \alpha \left(\frac{d - 1}{2n} + o \left(\frac{1}{n} \right) \right) + (1 - \alpha) \Delta.
\end{IEEEeqnarray}    
Picking $\alpha$ to be $0$ or $1$ depending on the relative values of the two terms gives the desired result. 

\subsection{Proof of Theorem~\ref{thm:dist_est_lower_bound}}

First, note that 
\begin{IEEEeqnarray}{rCl}
    \IEEEeqnarraymulticol{3}{l}{\min_{\hat \pi} \max_{\pi \in B_{\text{KL}}(\pitil, \Delta)} \EE[\dKL(\pi \| \hat \pi)]} \nonumber \\ 
    \qquad \qquad &\overset{(a)}{\ge} & \min_{\hat \pi} \max_{\pi \in B_{\text{KL}}(\pitil, \Delta)} \frac{\EE [\|\pi - \hat \pi\|_1^2 ]}{2} \nonumber\\
    &\overset{(b)}{\ge}& \min_{\hat \pi} \max_{\pi \in B_{\text{KL}}(\pitil, \Delta)} \frac{\EE [\|\pi - \hat \pi\|_1]^2}{2} \label{eqn:pinsker_lb},
\end{IEEEeqnarray}
where $(a)$ follows from Pinsker's inequality and $(b)$ follows using $\EE[X^2] \ge \EE[X]^2$. In the rest of the proof, we lower bound the $\ell_1$ norm error using Assouad's lemma \cite{bin_yu} and use~\eqref{eqn:pinsker_lb} to get a lower bound on the KL--divergence.
\begin{lemma}[Assouad's Lemma]
\label{lem:assouad}
Consider distributions $\theta_v$ on $d$ letters, where $v \in \{-1, 1\}^l$, such that
\begin{equation}
    \label{eqn:condition_assouad}
    \min_{a \in \mathcal{A}} \left(L(\theta_v, a) + L(\theta_{v'}, a)\right) \ge \lambda \cdot d_H(v, v'),
\end{equation}   
where $\mathcal{A}$ is a family of distributions on $d$ letters, $L(\cdot,\cdot)$ is a real-valued loss function defined between distributions, $d_H(\cdot,\cdot)$ denotes Hamming distance between vectors and $\lambda$ is a constant. Then,
\begin{equation}
    \min_{\hat \theta} \max_{\theta \in \mathcal{A}} L(\theta, \hat \theta) \ge \frac{l \lambda}{4} \exp(- n \max\limits_{v, v': d_H(v, v') = 1} \dKL(\theta_v \| \theta_{v'})).
\end{equation}
\end{lemma}

Without loss of generality, assume that $d$ is even. We take $l = \frac {d}2$, $\mathcal{A} = B_{\text{KL}}(\pitil, \Delta)$ for $\pitil = \Unif[d]$, and the loss $L$ to be the expected $\ell_1$--norm. Clearly, $v \in \{-1, +1\}^{\frac {d}2}$. Define $\theta_v$ as follows:
\begin{equation}
    \theta_v = \left(\frac{1}{d} + \tau v_1, \frac{1}{d} - \tau v_1, \dots, \frac{1}{d} + \tau v_{\frac d2}, \frac{1}{d} - \tau v_{\frac {d}2} \right).
\end{equation}

\begin{enumerate}
    \item To ensure non-negativity of coordinates, we need $\tau \le \frac 1{d}$. We take $\tau \le \frac{1}{2 d}$ to ensure each $\frac{1}{d} - \tau \ge \frac {1}{2 d}$. It can be seen that $\theta_v$ is a valid distribution for $d \ge 2$.
    
    \item To ensure that $\theta_v \in B_{\text{KL}}(\pitil, \Delta)$, note that we need $\dKL(\theta_v \| \pitil) \le \Delta$. The sum of the $(2i - 1)$-th term and the $(2i)$-th term in the KL--divergence is
    \begin{multline}
        \left(\frac{1}{d} + \tau v_i \right) \log \left(\frac{\frac{1}{d} + \tau v_i}{\frac{1}{d}} \right) \\ + \left(\frac{1}{d} - \tau v_i \right) \log \left(\frac{\frac{1}{d} - \tau v_i}{\frac{1}{d}} \right).
    \end{multline}
    Since $\tau \le \frac {1}{2d}$, by a Taylor expansion of this function it can be seen that this term is at most \begin{equation}
        \frac{\tau^2}{2} \left(\frac{1}{\frac{1}{d} - \tau} + \frac{1}{\frac{1}{d} - \tau} \right) \le \frac{\tau^2}{2} 4 d.
    \end{equation}
    Summing this up over the $\frac{d}{2}$ such pairs, we see that the $\dKL(\theta_v \| \pitil) \le d^2 \tau^2$.  To ensure this is at most $\Delta$, we need $\tau \le \sqrt{\Delta/d^2}$.
        
    \item To ensure that \eqref{eqn:condition_assouad} is satisfied in Lemma~\ref{lem:assouad}, we need
    \begin{IEEEeqnarray}{rCl}
        \min_{a \in \mathcal{A}} \| \theta_v - a \|_1 + \| a - \theta_v' \|_1
         & \ge & \frac{\| \theta_v - \theta_v' \|_1}{2} \nonumber\\
         & = & \frac{d_H (v, v')}{2} (4 \tau). \nonumber    \end{IEEEeqnarray}
    It can be seen that \eqref{eqn:condition_assouad} holds for $\lambda = 2 \tau$.

    \item Next, we compute 
        \begin{equation}\max_{v, v': d_H(v, v') = 1} \dKL(\theta_v \| \theta_{v'}).
    \end{equation}
    Since the Hamming distance is at most $1$, $v$ and $v'$ differ only at one coordinate, say $i$. Thus, $\theta_v$ and $\theta_{v'}$ would be different only at coordinates $2 i - 1$ and $2 i$.  
    Note that
    \begin{IEEEeqnarray}{rCl}
    \dKL(\theta_v \| \theta_{v'}) &\le& \left(\frac{1}{d} + \tau v_i \right) \log \left(\frac{\frac{1}{d} + \tau v_i}{\frac{1}{d} - \tau v_i} \right) \nonumber\\
    & & +\left( \frac{1}{d} - \tau v_i \right) \log \left(\frac{\frac{1}{d} - \tau v_i}{\frac{1}{d} + \tau v_i} \right)\nonumber \\
    &\le& \frac{(2 \tau)^2}{\frac{1}{2d}} \le 8 \tau^2 d.
\end{IEEEeqnarray}
    \end{enumerate}
Using Assouad's Lemma gives a lower bound of 
\begin{equation}
    \min_{\hat \pi} \max_{\pi \in B_{\text{KL}}(\pitil, \Delta)} \EE [\|\pi - \hat \pi\|_1] \ge \frac 14 \, \frac {d}{2} \, 2 \tau \, e^{-n 8 \tau^2 d}.
\end{equation}
To ensure that the term in the exponent is $O(1)$, we need for instance $n 8 \tau^2 d \le 2$, i.e. $\tau \le \frac{1}{\sqrt{4 dn}}$. Note that to satisfy the first two constraints, $\tau \le \min \left(\frac 1{2d}, \frac{\sqrt{\Delta}}{d}\right)$.

Pick $\tau = \min \left(\frac{1}{\sqrt{4 dn}}, \frac 1{2d}, \frac{\sqrt{\Delta}}{d}\right)$. This gives a lower bound of
\begin{equation}
    \frac {1}{4} \min \left(\sqrt{\frac{d}{4 n}}, \frac 1{2}, \sqrt{\Delta} \right) e^{-2}.
\end{equation}
This is a lower bound on the expected $\ell_1$ norm. To get a lower bound on the KL--divergence itself, we use~\eqref{eqn:pinsker_lb}. 

\subsection{Proof of Theorem~\ref{thm:dist_est_upper_bound_synonym_sample}}

First, note that
\begin{equation}
    \EE[\dKL(\pi \| \hat \pi)] \le \alpha \EE[\dKL(\pi \| \piopt)] + (1 - \alpha) \EE[\dKL(\pi \| \pitilopt)]
\label{eqn:alpha_interpolation}
\end{equation}
which follows from the convexity of KL divergence.

Note from Theorem~\ref{thm:dist_est_upper_bound_optimal} that $\EE[\dKL(\pi \| \piopt)] \le \frac{d - 1}{2 n} + o \left(\frac{d}{n} \right)$. Note for the other term that
\begin{IEEEeqnarray}{rCl}
    \IEEEeqnarraymulticol{3}{l}{\dKL(\pi \| \pitilopt)} \nonumber \\
    &=& \sum_{i \in [d_0]} \pi_i \log\left(\frac{\pi_i}{\pitilopt(i)}\right)  \nonumber\\
    &=& \sum_{i \in [d_0]} \left(\pi_i \log\left(\frac{\pi_i}{\pitil_i}\right) + \pi_i \log\left(\frac{\pitil_i}{\pitilopt(i)} \right)\right).
\end{IEEEeqnarray}
The first term is $\dKL(\pi, \pitil) \le \Delta$. We are left to bound the expected value of the second term. Note that for symbol counts $N_1, N_2, \dots, N_{d_0}$ from the samples $Y^{n_0}$, $\pitilopt(i) = \frac{N_i + \beta_i}{n_0 + \gamma}$, where $\gamma = \sum_{i \in [d_0]} \beta_i$ for suitable $\beta_i$ from~\cite{kl_div_estimation_3}. We have 
\begin{IEEEeqnarray}{rCl}
\IEEEeqnarraymulticol{3}{l}{\EE \left[ \log\left(\frac{\pitil_i}{\pitilopt(i)} \right) \right] } \nonumber\\
& = & \EE \left[ \log\left(\frac{\pitil_i (n_0 + \gamma)}{N_i + \beta_i} \right) \right] \nonumber\\
& = & \log \left(1 + \frac{\gamma}{n_0} \right) + \EE \left[ \log\left(\frac{\pitil_i n_0 }{N_i + \beta_i} \right) \right] \nonumber \\
& \overset{(a)}{\le} & \log \left(1 + \frac{\gamma}{n_0} \right) + \EE \left[ \log\left(\frac{\pitil_i n_0 }{(N_i + 1) \min(\beta_i, 1)} \right) \right]
\end{IEEEeqnarray}
\begin{IEEEeqnarray}{rCl}
& = & \log \left(1 + \frac{\gamma}{n_0} \right) - \log(\min(\beta_i, 1)) + \EE \left[ \log\left(\frac{\pitil_i n_0 }{N_i + 1} \right) \right] \nonumber \\
& = & \log \left(1 + \frac{\gamma}{n_0} \right) + \log \left( \frac{1}{\beta_i} \right)_+ + \EE \left[ \log\left(\frac{\pitil_i n_0 }{N_i + 1} \right) \right] \nonumber \\
& \overset{(b)}{\le} & \log \left(1 + \frac{d_0}{n_0} \right) + \log \left(2\right) + \EE \left[ \log\left(\frac{\pitil_i n_0 }{N_i + 1} \right) \right],
\label{eqn:expectation_pi0}
\end{IEEEeqnarray}
where $(a)$ uses $N_i + \beta_i \ge (N_i + 1) \min(\beta_i, 1)$, and $(b)$ uses $\beta_i \ge \frac 12$ and $\gamma \le d_0$ for the optimal estimator from \cite{kl_div_estimation_3}.

We will finally prove $\EE \left[ \log\left(\frac{\pitil_i n_0 }{N_i + 1} \right) \right] \le 0$.
\begin{IEEEeqnarray}{rCl}
\EE \left[ \log\left(\frac{\pitil_i n_0 }{N_i + 1} \right) \right] & \overset{(a)}{\le} & \EE \left[ \left(\frac{\pitil_i n_0 }{N_i + 1} - 1 \right) \right] \\
& \overset{(b)}{\le} & \left(\frac{\pitil_i n_0}{(n_0+1)\pitil_i} - 1 \right) \le 0,
\end{IEEEeqnarray}
where $(a)$ follows since $\log(x) \le x - 1$ for $x > 0$, and $(b)$ uses the standard result that $\EE \left[\frac {1}{N_i + 1} \right] \le \frac{1}{(n_0+1)\pitil_i}$. So, we have
\begin{equation}
    \EE \left[ \log\left(\frac{\pitil_i}{\pitilopt(i)} \right) \right] \le \log \left(1 + \frac{d_0}{n_0} \right) + \log \left(2\right).
\end{equation}
Multiplying the above by $\pi_i$, summing over all $i$ and picking $\alpha$ to be $0$ or $1$ in \eqref{eqn:alpha_interpolation} appropriately gives us the desired result. 

\bibliographystyle{IEEEtran}
\bibliography{refs}

@inproceedings{xu2023why,
  title     = {Why Do Nearest Neighbor Language Models Work?},
  author    = {Xu, Frank F. and Alon, Uri and Neubig, Graham},
  booktitle = {Proceedings of the 40th International Conference on Machine Learning},
  series    = {Proceedings of Machine Learning Research},
  volume    = {202},
  pages     = {38325--38341},
  year      = {2023},
  publisher = {PMLR}
}

@inproceedings{khandelwal2020generalization,
  title     = {Generalization through Memorization: Nearest Neighbor Language Models},
  author    = {Khandelwal, Urvashi and Levy, Omer and Jurafsky, Dan and Zettlemoyer, Luke and Lewis, Mike},
  booktitle = {International Conference on Learning Representations},
  year      = {2020}
}

@inproceedings{malagutti2024role,
  title     = {The Role of {$n$}-gram Smoothing in the Age of Neural Networks},
  author    = {Malagutti, Luca and Buinovskij, Andrius and Svete, Anej and Meister, Clara and Amini, Afra and Cotterell, Ryan},
  booktitle = {Proceedings of the 2024 Conference of the North American Chapter of the Association for Computational Linguistics: Human Language Technologies},
  pages     = {6882--6899},
  year      = {2024},
  address   = {Mexico City, Mexico},
  publisher = {Association for Computational Linguistics}
}

@inproceedings{falahatgar_20,
  title={Towards Competitive N-gram Smoothing},
  author={Falahatgar, Moein and Ohannessian, Mesrob and Orlitsky, Alon and Pichapati, Venkatadheeraj},
  booktitle={International Conference on Artificial Intelligence and Statistics},
  pages={4206--4215},
  year={2020},
  organization={PMLR}
}

@book{elements_of_information_theory,
  author = {Cover, Thomas M. and Thomas, Joy A.},
  howpublished = {Hardcover},
  isbn = {0471241954},
  publisher = {Wiley-Interscience},
  title = {Elements of Information Theory 2nd Edition},
  year = {2006}
}

@INPROCEEDINGS{llm_alignment,
  author={Yang, Joy Qiping and Salamatian, Salman and Sun, Ziteng and Suresh, Ananda Theertha and Beirami, Ahmad},
  booktitle={2024 IEEE International Symposium on Information Theory (ISIT)}, 
  title={Asymptotics of Language Model Alignment}, 
  year={2024},
  volume={},
  number={},
  pages={2027-2032},
  doi={10.1109/ISIT57864.2024.10619456}}

@InProceedings{kl_div_estimation_1,
author="Braess, Dietrich
and Forster, J{\"u}rgen
and Sauer, Tomas
and Simon, Hans U.",
editor="Cesa-Bianchi, Nicol{\`o}
and Numao, Masayuki
and Reischuk, R{\"u}diger",
title="How to Achieve Minimax Expected Kullback-Leibler Distance from an Unknown Finite Distribution",
booktitle="Algorithmic Learning Theory",
year="2002",
publisher="Springer Berlin Heidelberg",
address="Berlin, Heidelberg",
pages="380--394",
isbn="978-3-540-36169-5"
}

@inproceedings{kl_div_estimation_2,
 author = {Paninski, Liam},
 booktitle = {Advances in Neural Information Processing Systems},
 editor = {L. Saul and Y. Weiss and L. Bottou},
 pages = {},
 publisher = {MIT Press},
 title = {Variational Minimax Estimation of Discrete Distributions under KL Loss},
 volume = {17},
 year = {2004}
}

@article{kl_div_estimation_3,
title = {Bernstein polynomials and learning theory},
journal = {Journal of Approximation Theory},
volume = {128},
number = {2},
pages = {187-206},
year = {2004},
issn = {0021-9045},
doi = {https://doi.org/10.1016/j.jat.2004.04.010},
url = {https://www.sciencedirect.com/science/article/pii/S0021904504000723},
author = {Dietrich Braess and Thomas Sauer}
}

@Inbook{bin_yu,
author="Yu, Bin",
editor="Pollard, David
and Torgersen, Erik
and Yang, Grace L.",
title="Assouad, Fano, and Le Cam",
bookTitle="Festschrift for Lucien Le Cam: Research Papers in Probability and Statistics",
year="1997",
publisher="Springer New York",
address="New York, NY",
pages="423--435",
isbn="978-1-4612-1880-7"
}

@misc{word2vec,
      title={Efficient Estimation of Word Representations in Vector Space}, 
      author={Tomas Mikolov and Kai Chen and Greg Corrado and Jeffrey Dean},
      year={2013},
      eprint={1301.3781},
      archivePrefix={arXiv},
      primaryClass={cs.CL},
      url={https://arxiv.org/abs/1301.3781}, 
}

@ARTICLE{cover_dist_est,
  author={Cover, T.},
  journal={IEEE Transactions on Information Theory}, 
  title={Admissibility properties or {G}ilbert's encoding for unknown source probabilities (Corresp.)}, 
  year={1972},
  volume={18},
  number={1},
  pages={216-217},
  doi={10.1109/TIT.1972.1054738}}

@ARTICLE{KT_dist_est,
  author={Krichevsky, R. and Trofimov, V.},
  journal={IEEE Transactions on Information Theory}, 
  title={The performance of universal encoding}, 
  year={1981},
  volume={27},
  number={2},
  pages={199-207},
  doi={10.1109/TIT.1981.1056331}}

@techreport{catoni_laplace_est,
      author        = "Catoni, O",
      title         = "{The Mixture Approach To Universal Model Selection}",
      year          = "1997",
      url           = "https://cds.cern.ch/record/461892",
}

@misc{ISIT_paper,
      title={Distribution Estimation with Side Information}, 
      author={Haricharan Balasundaram and Andrew Thangaraj},
      year={2026},
      eprint={2601.08535},
      archivePrefix={arXiv},
      primaryClass={cs.IT},
      url={https://arxiv.org/abs/2601.08535}, 
}

@inproceedings{chen_goodman,
    title = "An Empirical Study of Smoothing Techniques for Language Modeling",
    author = "Chen, Stanley F.  and
      Goodman, Joshua",
    booktitle = "34th Annual Meeting of the Association for Computational Linguistics",
    month = jun,
    year = "1996",
    address = "Santa Cruz, California, USA",
    publisher = "Association for Computational Linguistics",
    url = "https://aclanthology.org/P96-1041/",
    doi = "10.3115/981863.981904",
    pages = "310--318"
}

@article{lidstone,
  title   = {Note on the General Case of the {B}ayes-{L}aplace Formula for Inductive or a Posteriori Probabilities},
  author  = {Lidstone, George James},
  journal = {Transactions of the Faculty of Actuaries},
  volume  = {8},
  year    = {1920},
  pages   = {182--192}
}

@inproceedings{jelenik_mercer,
  title     = {Interpolated estimation of {M}arkov source parameters from sparse data},
  author    = {Jelinek, Frederick and Mercer, Robert L.},
  booktitle = {Proceedings of the Workshop on Pattern Recognition in Practice},
  year      = {1980},
  address   = {Amsterdam, The Netherlands},
  publisher = {North-Holland},
  month     = {May}
}

@article{katz,
  title   = {Estimation of probabilities from sparse data for the language model component of a speech recognizer},
  author  = {Katz, Slava M.},
  journal = {IEEE Transactions on Acoustics, Speech, and Signal Processing},
  volume  = {35},
  number  = {3},
  pages   = {400--401},
  year    = {1987},
  month   = {March}
}

@inproceedings{kneser_ney,
  title     = {Improved backing-off for m-gram language modeling},
  author    = {Kneser, Reinhard and Ney, Hermann},
  booktitle = {Proceedings of the IEEE International Conference on Acoustics, Speech, and Signal Processing (ICASSP)},
  volume    = {1},
  pages     = {181--184},
  year      = {1995}
}

@inproceedings{attention_is_all_you_need,
 author = {Vaswani, Ashish and Shazeer, Noam and Parmar, Niki and Uszkoreit, Jakob and Jones, Llion and Gomez, Aidan N and Kaiser, \L ukasz and Polosukhin, Illia},
 booktitle = {Advances in Neural Information Processing Systems},
 editor = {I. Guyon and U. Von Luxburg and S. Bengio and H. Wallach and R. Fergus and S. Vishwanathan and R. Garnett},
 pages = {},
 publisher = {Curran Associates, Inc.},
 title = {Attention is All you Need},
 volume = {30},
 year = {2017}
}

@inproceedings{glove,
    title = "{G}lo{V}e: Global Vectors for Word Representation",
    author = "Pennington, Jeffrey  and
      Socher, Richard  and
      Manning, Christopher",
    editor = "Moschitti, Alessandro  and
      Pang, Bo  and
      Daelemans, Walter",
    booktitle = "Proceedings of the 2014 Conference on Empirical Methods in Natural Language Processing ({EMNLP})",
    month = oct,
    year = "2014",
    address = "Doha, Qatar",
    publisher = "Association for Computational Linguistics",
    url = "https://aclanthology.org/D14-1162/",
    doi = "10.3115/v1/D14-1162",
    pages = "1532--1543"
}

@inproceedings{word2vec_2,
 author = {Mikolov, Tomas and Sutskever, Ilya and Chen, Kai and Corrado, Greg S and Dean, Jeff},
 booktitle = {Advances in Neural Information Processing Systems},
 editor = {C.J. Burges and L. Bottou and M. Welling and Z. Ghahramani and K.Q. Weinberger},
 pages = {},
 publisher = {Curran Associates, Inc.},
 title = {Distributed Representations of Words and Phrases and their Compositionality},
 url = {https://proceedings.neurips.cc/paper_files/paper/2013/file/9aa42b31882ec039965f3c4923ce901b-Paper.pdf},
 volume = {26},
 year = {2013}
}

@inproceedings{kim2021lipschitz,
  author    = {Kim, Hyunjik and Papamakarios, George and Mnih, Andriy},
  title     = {The {L}ipschitz Constant of Self-Attention},
  booktitle = {Proceedings of the 38th International Conference on Machine Learning (ICML)},
  year      = {2021},
  pages     = {5562--5571},
  publisher = {PMLR},
  series    = {Proceedings of Machine Learning Research},
  volume    = {139},
  url       = {https://proceedings.mlr.press/v139/kim21i.html}
}

@inproceedings{castin2024smooth,
  author    = {Castin, Val{\'e}rie and Ablin, Pierre and Peyr{\'e}, Gabriel},
  title     = {How Smooth Is Attention?},
  booktitle = {Proceedings of the 41st International Conference on Machine Learning},
  series    = {Proceedings of Machine Learning Research},
  volume    = {235},
  pages     = {5817--5840},
  year      = {2024},
  publisher = {PMLR},
  month     = {Jul},
  url       = {https://proceedings.mlr.press/v235/castin24a.html}
}

@inproceedings{ye2023lipschitz,
  author    = {Ye, Mao and Liu, Yulong and Chen, Zhenyu and Wang, Shiyu},
  title     = {Mitigating Transformer Overconfidence via {L}ipschitz Regularization},
  booktitle = {Proceedings of the Thirty-Ninth Conference on Uncertainty in Artificial Intelligence},
  series    = {Proceedings of Machine Learning Research},
  volume    = {216},
  pages     = {2422--2432},
  year      = {2023},
  publisher = {PMLR},
  url       = {https://proceedings.mlr.press/v216/ye23a.html}
}

@misc{wu2015chebyshev,
  title  = {Chebyshev Polynomials, Moment Matching, and Optimal Estimation of the Unseen},
  author = {Wu, Yihong and Yang, Pengkun},
  year   = {2015},
  eprint = {1504.01227},
  archivePrefix = {arXiv},
  primaryClass  = {cs.IT}
}

@article{radford2018improving,
  title   = {Improving Language Understanding by Generative Pre-Training},
  author  = {Radford, Alec and Narasimhan, Karthik and Salimans, Tim and Sutskever, Ilya},
  year    = {2018},
  journal = {OpenAI},
  url     = {https://cdn.openai.com/research-covers/language-unsupervised/language_understanding_paper.pdf}
}

@article{merity2016pointer,
  title   = {Pointer Sentinel Mixture Models},
  author  = {Merity, Stephen and Xiong, Caiming and Bradbury, James and Socher, Richard},
  journal = {arXiv preprint arXiv:1609.07843},
  year    = {2016},
  url     = {https://arxiv.org/abs/1609.07843}
}

@inproceedings{lhoest2021datasets,
  title     = {Datasets: A Community Library for Natural Language Processing},
  author    = {Lhoest, Quentin and Villanova del Moral, Albert and Jernite, Yacine and others},
  booktitle = {Proceedings of EMNLP: System Demonstrations},
  year      = {2021},
  pages     = {175--184},
  url       = {https://aclanthology.org/2021.emnlp-demo.21/}
}

\appendices

\section{Common Embeddings}
\label{appendix:embeddings}

Many models for language used while training embeddings satisfy the $(L, \epsilon)$-Lipschitz-logit process defined in \eqref{eqn:softmax_probab}.

\subsection{Word2Vec}

Let $e_w \in \mathbb{R}^t$ denote the Word2Vec `input' embedding of each word $w$. Note that $E_c \in \RR^{t \times |c|}$ is the input context embedding matrix. Under the Continuous Bag of Words (CBOW) model, this is converted to the final context embedding $f_c\in \mathbb{R}^t$ by averaging the individual word embeddings, i.e. \begin{equation}
    f_c = \frac 1k E_c 1_{|c| \times 1}.
\end{equation}
where $1_{|c| \times 1}$ denotes the $|c| \times 1$ matrix of all ones. This process is Lipschitz with $L_f = \frac{\sqrt{|c|}}{k}$.

The probability estimates in the model are trained to be
\begin{equation}
    \PP(w | c) = \frac{\exp(f_c^T \tilde e_w+ \epsilon_{c, w})}{\sum_{y \in V} \exp(f_c^T \tilde e_y+ \epsilon_{c, y})},
\end{equation}
where $\tilde e_w$'s are `output' embeddings of $w$ and $\epsilon_{c, w}$ includes possible training error in the model. The function $g_w(x) = x' \tilde e_w$ is $L$-Lipschitz for $L = \max_{w \in V} \| \tilde e_w \|_2$.

\subsection{Transformer-based embeddings}

A transformer takes the embedding context $E_c$ from the word embeddings $e_w$ and first converts it to $y_{t \times |c|}$ using the attention mechanism \cite{attention_is_all_you_need}. Recall that the last context position, $y_{|c|}$, is taken to be $f_c$. As before, the probabilities in the model are trained to be
\begin{equation}
    \PP(w | c) = \frac{\exp(f_c' e_w + \epsilon_{c, w})}{\sum_{y \in V} \exp(f_c' e_y + \epsilon_{c, y})},
\end{equation}
where $\epsilon_{c, w}$ includes possible training errors. 
The function $g_w(x) = x' e_w$ is $L$-Lipschitz for $L = \max_{w \in V} \| e_w \|_2$ and taking $f_c(y_{t \times |c|}) = y_{|c|}$ is $1$-Lipschitz.

It is left to show that the conversion of $E_c$ to $y_{t \times |c|}$ is $L_f$ Lipschitz. A transformer consists of $\ell$ sequential layers. If we prove that the $i$th layer is $L_i$-Lipschitz, then we get $L_f = \prod_{i = 1}^\ell L_i$. So, it is sufficient to show that the individual layers are Lipschitz. 

\noindent\textbf{Self Attention}: Self-attention is computed as \cite{attention_is_all_you_need}
\begin{equation}
        f(X) = \text{softmax}\left(\frac{QK^\top}{\sqrt{d}}\right)V,        
\end{equation}
where $Q = XW_Q$, $K = XW_K$, $V = XW_V$.
Without any input constraints, the self-attention step is not Lipschitz \cite{kim2021lipschitz}. However, as shown in \cite[Theorem 3.3]{castin2024smooth}, when the norm of the input vectors is bounded, the Lipschitz constant of self-attention scales as $\sqrt{n}$ for a sequence length $n$. From \cite[Lemma 3.8]{castin2024smooth}, note that if each head is Lipschitz and the matrices have finite norm, the overall multi-headed attention block is also Lipschitz.

\noindent\textbf{Residual Connections}: A layer of the form $f(x) = x + h(x)$ is $(1 + L)$-Lipschitz if $h$ is $L$-Lipschitz. Since $h$ is either a linear or a self attention layer and both of them are Lipschitz, $f$ is Lipschitz.
    
\noindent\textbf{Layer Normalization}: The LayerNorm is expressed as:
\begin{equation}
    f(x) = \frac{x - \mu(x)}{\sqrt{\sigma^2(x) + \epsilon}} \, \gamma + \beta,  
\end{equation}
where $\gamma,\beta$ are parameters, $\mu(x)=\frac1d\sum_{i=1}^d x_i$, $\sigma^2(x)=\frac1d\sum_{i=1}^d (x_i-\mu(x))^2$, and $\epsilon>0$ is a small constant added for numerical stability. It is shown in \cite[Supplementary material, Section A]{ye2023lipschitz} that the LayerNorm is  Lipschitz.

Transformer components such as MLP/feed-forward block and activation functions are known to be Lipschitz.

\subsection{GloVe}
    
GloVe embeddings \cite{glove} are trained to ensure the following:
\begin{equation}
    \log(X_{\tilde{w}w}) = f'_{\tilde{w}} e_w + b_{\tilde{w}} + \tilde{b}_w + \epsilon_{\tilde{w}, w},
    \label{eqn:glove_embedding}
\end{equation}
where $X_{\tilde{w}w}$ is the number of times word $w$ occurs as the word to be predicted when word $\tilde{w}$ is in the context $c$, $f_{\tilde{w}}$ is the embedding of the word $\tilde{w}$ when it acts as the context, $e_w$ is the embedding of the $w$ when it acts as the word to be predicted, and $\epsilon_{\tilde{w}, w}$ is the training error.

For a bigram model, the context is the previous word in the GloVe model, i.e. $c = \tilde{w}$. In such a case, the context embedding $f_c = e_{\tilde{w}}$. If sufficiently many samples are taken, due to stationarity and ergodicity, note that
\begin{equation}
    \frac{X_{cw}}{X_c} \approx \PP(W_t = w | W_{t-1} = c),
    \label{eqn:markov_stationarity}
\end{equation}
where $X_c = \sum_{y \in V_c} X_{cy}$.

Combining equations~\eqref{eqn:glove_embedding} and~\eqref{eqn:markov_stationarity}, we get 
\begin{IEEEeqnarray}{rCl}
    \PP(w | c) & \propto & \exp(f'_c e_w + b_c + \tilde{b}_w + \epsilon_{c, w}) \\
    & \propto & \exp(f'_c e_w + \tilde{b}_w + \epsilon_{c, w}).
\end{IEEEeqnarray}
From the above, we observe that the next-word probabilities are Lipschitz-logit with $g_w(x) = x' e_w + \tilde{b}_w$, and $g_w$ is Lipschitz with a constant $\sup_{w} \| e_w \|$.

\end{document}